\documentclass{aa}

\usepackage{enumitem}
\usepackage{graphicx}
\usepackage{txfonts}
\usepackage{natbib}
\bibpunct{(}{)}{;}{a}{}{,}

\usepackage{lscape}
\usepackage{longtable}
\usepackage{amssymb}

\def\G23{G023.01$-$00.41}

\usepackage{color}

\begin{document}

  \title{Physical conditions in the warped accretion disk of a massive star}
  \subtitle{349\,GHz ALMA observations of G023.01$-$00.41}

  \author{A.~Sanna\inst{1, 2, 3} \and A.~Giannetti\inst{2} \and M.~Bonfand\inst{4}
  \and L.~Moscadelli\inst{5} \and R.~Kuiper\inst{6}  \and J.~Brand\inst{2} \and R.~Cesaroni\inst{5}
   \and A.~Caratti\,o\,Garatti\inst{7} \and T.~Pillai\inst{8} \and K.\,M.~Menten\inst{3}}

                                      
   \institute{INAF, Osservatorio Astronomico di Cagliari, via della Scienza 5, 09047, Selargius, Italy \\
   \email{alberto.sanna@inaf.it}
   \and INAF - Istituto di Radioastronomia \& Italian ALMA Regional Centre, Via P. Gobetti 101, I-40129 Bologna, Italy
   \and Max-Planck-Institut f\"{u}r Radioastronomie, Auf dem H\"{u}gel 69, 53121 Bonn, Germany
   \and Laboratoire d’astrophysique de Bordeaux, Universit\'{e} de Bordeaux, CNRS, B18N, all\'{e}e Geoffroy Saint-Hilaire, 33615 Pessac, France
   \and INAF, Osservatorio Astrofisico di Arcetri, Largo E. Fermi 5, 50125 Firenze, Italy   
  \and Institute of Astronomy and Astrophysics, University of T\"{u}bingen, Auf der Morgenstelle 10, D-72076 T\"{u}bingen, Germany   
  \and Dublin Institute for Advanced Studies, Astronomy \& Astrophysics Section, 31 Fitzwilliam Place, Dublin 2, Ireland
  \and Institute for Astrophysical Research, Boston University, 725 Commonwealth Ave, Boston, MA, 02215, USA
}  

   \date{Received 27 November 2020; Accepted 5 July 2021}


  \abstract{Young massive stars warm up the large amount of gas and dust which condenses in their vicinity, exciting a forest of lines from different 
  molecular species. Their line brightness is a diagnostic tool of the gas physical conditions locally, which we use to set constraints on the environment
  where massive stars form. We made use of the Atacama Large Millimeter/submillimeter Array at frequencies near 349\,GHz, with an angular resolution
  of $0\farcs1$, to observe the methyl cyanide (CH$_3$CN) emission which arises from the accretion disk of a young massive star. We sample the disk
  midplane with twelve distinct beams, where we get an independent measure of the gas (and dust) physical conditions. The accretion disk extends above the midplane
  showing a double-armed spiral morphology projected onto the plane of the sky, which we sample with ten additional beams: along these
  apparent spiral features, gas undergoes velocity gradients of about 1\,km\,s$^{-1}$ per 2000\,au. The gas temperature (T) rises symmetrically along each
  side of the disk, from about 98\,K at 3000\,au to 289\,K at 250\,au, following a power law with radius, R$^{-0.43}$. The CH$_3$CN column density (N)
  increases from 9.2\,$\times$\,10$^{15}$\,cm$^{-2}$ to 8.7\,$\times$\,10$^{17}$\,cm$^{-2}$ at the same radii, following a power
  law with radius, R$^{-1.8}$. In the framework of a circular gaseous disk observed approximately edge-on, we infer an  H$_2$ volume density
  in excess of 4.8\,$\times$\,$10^9$\,cm$^{-3}$ at a distance of 250\,au from the star. We study the disk stability against fragmentation following the
  methodology by \citet{Kratter2010}, appropriate under rapid accretion, and we show that the disk is marginally prone to fragmentation along its whole extent.}
  

   
   
   


   \maketitle
%

\section{Introduction}

The circumstellar regions within a few 1000\,au from early-type young stars, where the (average) kinetic temperature of the interstellar medium 
exceeds 100\,K, are rich in methyl cyanide gas (CH$_3$CN), whose relative abundance with respect to H$_2$ is typically larger than 10$^{-9}$
\citep[e.g.,][]{Hernandez2014}. CH$_3$CN is a symmetric top molecule whose millimeter spectrum is divided into groups of rotational
transitions, with the following favourable properties \citep[e.g.,][]{Boucher1980,Loren1984}: \textbf{(i)} each group corresponds to a single $J\to(J-1)$
transition with varying $K$, where $J$ and $K$ are the two quantum numbers of total angular momentum and its projection on the axis of molecular
symmetry, respectively; \textbf{(ii)} each group of transitions covers a ``narrow'' bandwidth, within about a GHz, and is separated by a few tens of GHz
from the lower and higher $J$ transitions;  \textbf{(iii)} the energy levels within a group are only populated through collisions, and their excitation
temperatures span many 100\,K; \textbf{(iv)} rotational transitions of the isotopologue CH$_3 ^{13}$CN emit within the same narrow bandwidth
of the CH$_3$CN transitions with same quantum numbers. These spectral properties make of CH$_3$CN emission a sensitive thermometer for the
interstellar gas, where the H$_2$ density exceeds a critical value of approximately $10^5$\,cm$^{-3}$ \citep[e.g.,][]{Shirley2015}; this critical density
would be reached, for instance, inside a spherical core of 0.1\,pc radius (R), gas mass of 50\,M$_{\odot}$, and density proportional to $\rm R^{-1.5}$.

Since the first CH$_3$CN interferometric observations by \citet{Cesaroni1994}, performed with the Plateau de Bure Interferometer at 110\,GHz,
CH$_3$CN emission and its isotopologue CH$_3 ^{13}$CN have been observed at (sub-)arcsecond resolution towards tens of early-type young
stars, in order to estimate the linewidth, temperature, and H$_2$ number density of local gas  \citep[e.g.,][]{Zhang1998,Beltran2005,Beuther2005,Chen2006,Qiu2011,Hernandez2014,Hunter2014,SanchezMonge2014,Zinchenko2015,
Bonfand2017,Ilee2018,Ahmadi2018,Johnston2020}. For instance, CH$_3$CN\,($12_{K}$--$11_{K}$) observations at 220\,GHz, whose lower excitation
energy (E$_{l}/k$) exceeds 58\,K, indicate average rotational (and kinetic) temperatures of 200\,K approximately, and CH$_3$CN column
densities of the order of 10$^{16}$\,cm$^{-2}$. These conditions are met at radii of 1000\,au from the brightness peak of the dust
continuum emission at the same frequency, whose position is assumed to pinpoint a young star.
Alternatively, temperature and density gradients measured through CH$_3$CN transitions are proxies for the radiative feedback of luminous stars in the
making ($>$\,$10^3$\,L$_{\odot}$).


\begin{table*}
\caption{Summary of ALMA observations at Cycle\,4 (project code 2016.1.01200.S).}\label{tobs}
\centering
\begin{tabular}{ c c c c c c c c c c}

\hline \hline
Array\,Conf. & R.A.\,(J2000) &        Dec.\,(J2000)    &   V$_{\rm LSR}$ & Freq.\,Cove. & $\Delta\nu$  &     BP\,Cal.        & Phase\,Cal. & Pol.\,Cal. &  HPBW   \\
                   &      (h\,m\,s)   &  ($^{\circ}$\,$'$\,$''$) &   (km\,s$^{-1}$) &     (GHz)       &     (kHz)     &                        &                   &               &  ($''$)     \\
    (1)          &          (2)        &               (3)              &            (4)           &        (5)        &      (6)        &       (7)            &      (8)        &      (9)    &    (10)    \\
\hline
 & & & & & & & & & \\
C40--6 & 18:34:40.290 & --09:00:38.30 &  77.4  & 335.6, 349.6 & 976.5 & J1751$+$0939 & J1825$-$0737 & J1733$-$1304 & 0.10 \\
\hline
\end{tabular}
\tablefoot{Column\,1: 12m-array configuration. Columns\,2 and\,3: target phase center (ICRS system). Column\,4: source radial velocity. Column\,5:
minimum and maximum rest frequencies covered with four basebands (BB1--4) and the LO frequency set at 343.5\,GHz. Column\,6: spectral resolution on
BB4. Columns\,7, 8, and\,9: bandpass, phase, and polarization (and absolute flux) calibrators employed. Calibration sources were set by the ALMA operators at the
time of the observations. Column\,10: beam size at a representative frequency of 338.57\,GHz.}
\end{table*}


In this paper, we report on spectroscopic CH$_3$CN, CH$_3$OH (methanol), and dust continuum observations with the Atacama Large Millimeter/submillimeter Array
(ALMA) at 349\,GHz with  an angular resolution of $0\farcs1$.  We exploit the CH$_3$CN\,($19_{K}$--$18_{K}$) K-ladder, with excitation energies ranging from
168\,K (for K\,$=$\,0) to 881\,K (for K\,$=$\,10), to probe, at different radii, the physical conditions in the accretion disk of an early-type young star. We targeted
the star-forming region G023.01$-$00.41, at a trigonometric distance of 4.59$^{+0.38}_{-0.33}$~kpc from the Sun \citep{Brunthaler2009}, where we recently
revealed the accretion disk around a young star of $10^{4.6}$\,L$_{\odot}$, corresponding to a ZAMS star of 20\,M$_{\odot}$ \citep[][their Fig.\,1]{Sanna2019};
the disk was imaged by means of spectroscopic ALMA observations of both CH$_3$CN and CH$_3$OH lines at $0\farcs2$ resolution in the 230\,GHz band. The disk
extends up to radii of 3000\,au from the central star where it warps above the midplane; here, we resolve the outer disk regions in two apparent spirals projected onto
the plane of the sky. We showed that
molecular gas is falling in and slowly rotating with sub-Keplerian  velocities down to radii of 500 au from the central star, where we measured a mass infall rate of
$6\times10^{-4}$\,M$_{\odot}$\,yr$^{-1}$ \citep[][their Fig.\,5]{Sanna2019}. The disk and star system drives a radio continuum jet and a molecular outflow aligned
along a position angle of $57\degr$, measured east of north \citep[][their Fig.\,2]{Sanna2016}; their projected axis is oriented perpendicular to the disk midplane whose
inclination with respect to the line-of-sight was estimated to be less than $30\degr$ (namely, the disk is seen approximately edge-on; \citealt{Sanna2014,Sanna2019}).
Previously, we also measured the average gas conditions over the same extent of the whole disk, by means of Submillimeter Array (SMA) observations of
the CH$_3$CN\,($12_{K}$--$11_{K}$) emission, and estimated a kinetic temperature of 195\,K and CH$_3$CN column density of $5.1\times10^{16}$\,cm$^{-2}$
\citep[][their Fig.\,2 and Table\,4]{Sanna2014}.



\begin{figure}
\centering
\includegraphics [angle= 0, scale= 0.8]{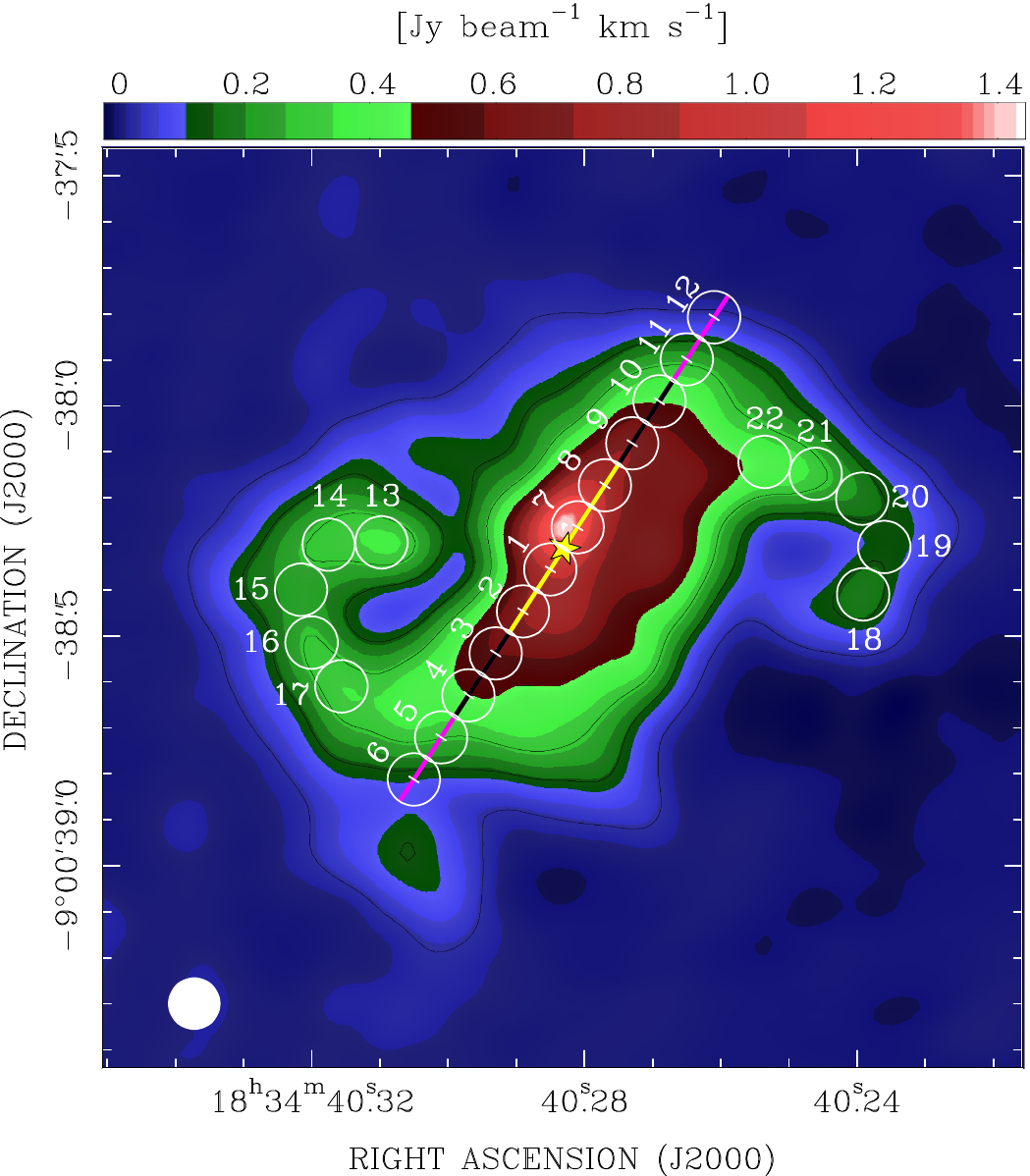}
\caption{Spatial morphology of dense gas around the most massive star (star symbol) in G023.01$-$00.41: moment-zero map of the
CH$_3$OH\,($14_{1,13}$--$14_{0,14}$)\,A$^+$ emission (colors). The upper wedge quantifies the line intensity from its peak to the maximum negative in the
map; black contours are drawn at levels of 10, 50, and 100 times the 1\,$\sigma$ rms of 2.8\,mJy\,beam$^{-1}$\,km\,s$^{-1}$.  The disk plane, with white ticks
at steps of 500\,au, is drawn from the central star position at a position angle of $-33\degr$ (see Sect.~\ref{res}). The synthesized ALMA beam is shown in the
bottom left corner.  White empty circles mark areas (hereafter, ``pointings'') where the CH$_3$CN spectra of Figs.\,\ref{figA3} and~\ref{figA4} are integrated
and are labeled from 1 to 22 (cf. column 1 of Table\,\ref{tpar}).}\label{fig1}
\end{figure}



\begin{figure*}
\centering
\includegraphics [angle= 0, scale= 0.9]{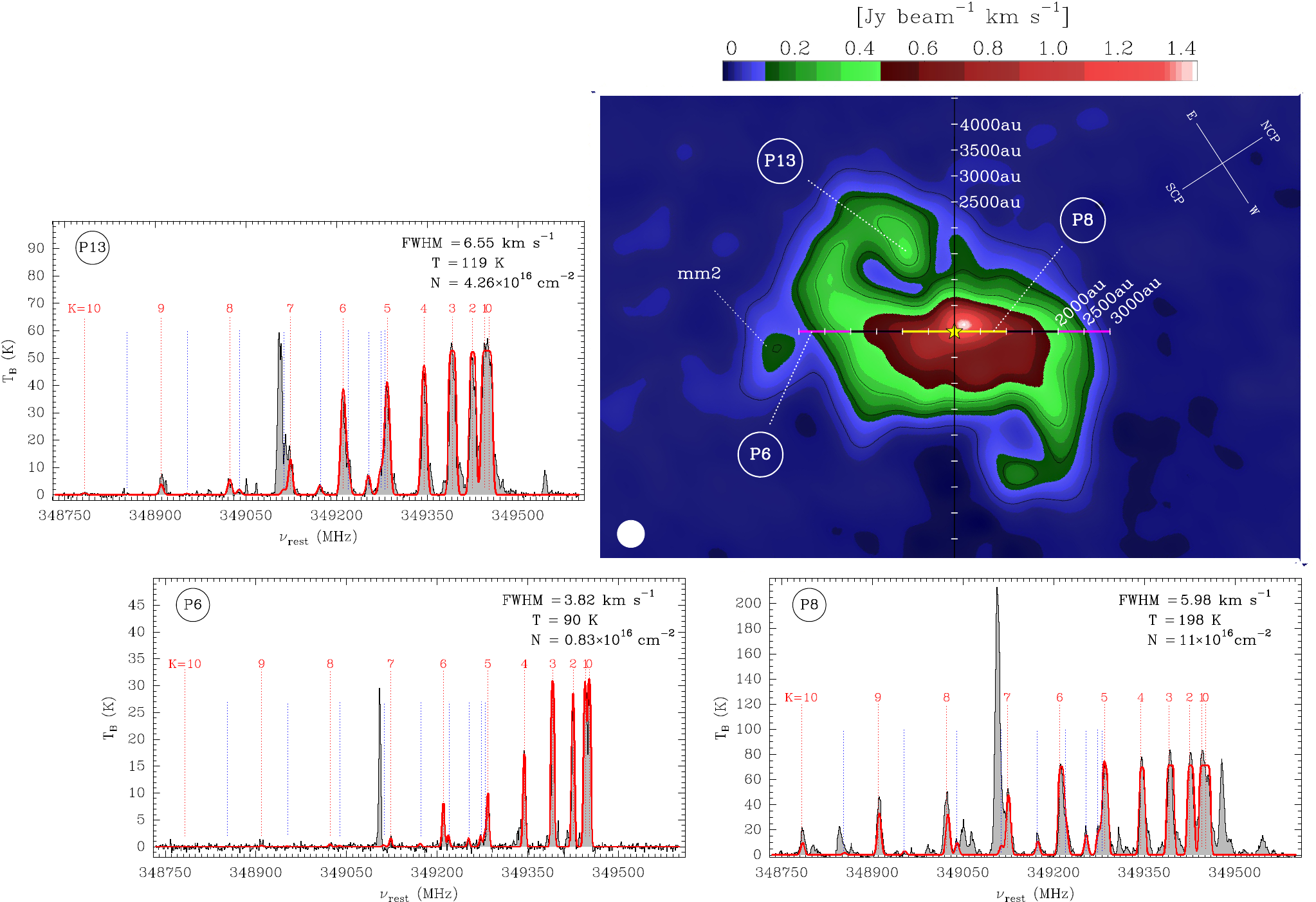}
\caption{Physical conditions in the accretion disk of G023.01$-$00.41. \textbf{Main panel:} same image (and symbols) as in Fig.\,\ref{fig1}, with the
equatorial reference system (top right) rotated clockwise by $57^{\circ}$, in order to align the (projected) outflow direction with the vertical axis of the plot.
For three selected pointings indicated in white, we plot the local integrated spectra at 349\,GHz in the side panels (grey histograms), as an example of the molecular
line profiles and fitting. The cluster source ``mm2'' is marked; this source might alter the accretion disk around the primary (star symbol).
\textbf{Side panels:} examples of different CH$_3$CN\,($19_{K}$--$18_{K}$) spectra extracted in the inner (P8), outer (P6), and apparent spiral regions (P13) of
the accretion disk (identifier on the top left). Positions of CH$_3$CN and CH$_3 ^{13}$CN components are labeled in red and blue, respectively, with K ranging
from 0 up to 10. The brightest line at a rest frequency of 349.107\,GHz corresponds to the CH$_3$OH emission imaged in the main panel; this line shows a bright
maser contribution in the inner-region spectra (e.g., P8). Red profiles draw the synthetic spectra fitted with MCWeeds by Monte Carlo Markov chains method
assuming LTE; best fit parameters of intrisic FWHM, excitation temperature, and column density of CH$_3$CN are listed on the top right (cf. Table\,\ref{tpar}).
We note that the excess of emission visible in the higher K lines of the observed spectrum at position P8, with respect to the synthetic spectrum, is likely due to
contamination from different molecular species.}
\label{fig2}
\end{figure*}


\section{Observations and calibration}\label{obs}

We observed the star-forming region G023.01$-$00.41 with the 12\,m-array of ALMA in band 7 (275--373\,GHz).
Observations were conducted under program 2016.1.01200.S on 2017 July 10 (Cycle\,4) during a 3\,hour run, with
precipitable water vapor of 0.35--0.38\,mm. The 12\,m-array observed with 40 antennas covering a baseline range
between 16\,m and 2647\,m, achieving an angular resolution and maximum recoverable scale of approximately
$0\farcs08$ and $1\farcs4$, respectively. Observation information is summarized in Table\,\ref{tobs}.

We made use of a mixed correlator setup consisting of four basebands (BB1--4), three operated in time division mode (TDM) and 
one in frequency division mode (FDM). Each TDM window had 64 spectral channels spaced over a bandwidth of 1875\,MHz. These bands were
used for continuum (full) polarization observations and tuned at the central frequencies of 336.57 (BB1), 338.57 (BB2), and 348.57\,GHz
(BB3) for optimal polarization performance (cf. ALMA Cycle 4 Proposer's Guide, Doc\,4.2. ver. 1.0, March 2016). The FDM window 
had 1920 spectral channels spaced over a bandwidth of 938\,MHz, to achieve a velocity resolution of  0.84\,km\,s$^{-1}$ after 
spectral averaging by a factor of 2. This band was used for spectral line observations and tuned at the central frequency of 349.15\,GHz (BB4),
and overlaps with the higher half of BB3. BB4 was placed to cover the K-ladder of the CH$_3$CN\,($19_{K}$--$18_{K}$) transition, with K
ranging from 0 to 10, and its isotopologue CH$_3 ^{13}$CN.

In the following, we report on the analysis of the high spectral-resolution band (BB4); the targeted molecular lines are listed in Table\,\ref{tlines} of 
the Appendix, and include a bright CH$_3$OH line previously identified in a preparatory SMA experiment (project code 2014B-S006).
The visibility data were calibrated with the Common Astronomy Software Applications (CASA) package, version\,4.7.2 (r39762), making
use of the calibration scripts provided with the quality assessment process (QA2). To determine the continuum level, we made use of the spectral image
cube and selected the line-free channels from a spectrum integrated over a circular area of $0\farcs5$ in size, which was centered on the target source.
The task \emph{uvcontsub} of CASA was used to subtract a constant continuum level across the spectral window (\emph{fitorder}\,$=$\,0). We imaged
the line and continuum emission with the task \emph{clean} of CASA setting a circular restoring beam size of $0\farcs114$, equal to the geometrical
average of the major and minor axes of the beam obtained with a Briggs's robustness parameter of 0.5. We achieve a sensitivity of
approximately 1\,mJy\,beam$^{-1}$ per resolution unit, which corresponds to a brightness temperature of about 1\,K over a beam of $0\farcs1$.
The continuum emission was integrated over a line free\footnote{The line free channels in the high spectral-resolution dataset were selected based on an
accurate eye inspection of the band, with the line frequencies of abundant molecular species marked in the spectrum to avoid them, and eventually comparing
maps produced with different selections of channels to exclude the effect of line contamination.} bandwidth of 52\,MHz achieving a sensitivity
of 0.2\,mJy\,beam$^{-1}$.

\begin{table*}
\caption{Physical parameters of the CH$_3$CN gas and dust observed near 349\,GHz towards the accretion disk in G023.01$-$00.41.}\label{tpar}
\centering
\begin{tabular}{ l r r r r c r c c r r r r}

\hline \hline
\multicolumn{1}{c}{ID}  &  \multicolumn{1}{c}{X-offset}  & \multicolumn{1}{c}{ Y-offset}  & \multicolumn{1}{c}{R$_{\rm p}$}  &  \multicolumn{1}{c}{S$_{\rm dust}$}  &
\multicolumn{1}{c}{v$_{\rm off}$}  & \multicolumn{1}{c}{FWHM$_{\rm int}$}& T$_{\rm rot}$  &  N$_{\rm CH_3CN}$ &\multicolumn{1}{c}{size} &
\multicolumn{1}{c}{$\tau_{4}$}  & \multicolumn{1}{c}{M$_{500}$} & \multicolumn{1}{c}{N$_{\rm H_2}$} \\
 & \multicolumn{1}{c}{($''$)} &  \multicolumn{1}{c}{($''$)} & \multicolumn{1}{c}{(au)} & \multicolumn{1}{c}{(mJy)} & (km\,s$^{-1}$) & \multicolumn{1}{c}{(km\,s$^{-1}$)} &
 (K) &  (10$^{16}$\,cm$^{-2}$) & \multicolumn{1}{c}{(mas)} & & (M$_{\odot}$)  &  (10$^{23}$\,cm$^{-2}$) \\
\multicolumn{1}{c}{(1)} & \multicolumn{1}{c}{(2)} & \multicolumn{1}{c}{(3)}  & \multicolumn{1}{c}{(4)}  & \multicolumn{1}{c}{(5)} & (6) & \multicolumn{1}{c}{(7)}  &
\multicolumn{1}{c}{(8)} & \multicolumn{1}{c}{(9)} & \multicolumn{1}{c}{(10)} & \multicolumn{1}{c}{(11)} & \multicolumn{1}{c}{(12)} & \multicolumn{1}{c}{(13)} \\
\hline
 & & & & & & & & & & & & \\
                   \multicolumn{13}{c}{\textbf{Eastern disk side}}        \\
P1    & --0.074 & --0.056 &   250 & 29.7 & +0.78$\pm$0.03 & 7.37$\pm$0.20 & $267^{+2}_{-2}$ &$56.75^{+0.53}_{-0.58}$ &  70 & 9.64    & 0.16 &10.0 \\
P2    & --0.015 & --0.147 &   750 & 13.6 & +3.25$\pm$0.03 & 5.17$\pm$0.56 & $197^{+2}_{-2}$ &$35.80^{+0.80}_{-0.73}$ &  95 & 15.00   & 0.10 &  6.2 \\
P3    & +0.045 & --0.238 & 1250 &   4.6 & +2.69$\pm$0.03 & 4.79$\pm$0.16 & $140^{+3}_{-2}$ &  $9.33^{+0.33}_{-0.31}$ &  95 & 6.31   & 0.05  &  2.9 \\
P4    & +0.104 & --0.330 & 1750 &   3.8 & +2.83$\pm$0.03 & 3.09$\pm$0.09 & $118^{+3}_{-2}$ &  $6.17^{+0.30}_{-0.27}$ &  90 & 7.01   & 0.05  &  2.9 \\
P5    & +0.163 & --0.421 & 2250 &   2.8 & +1.86$\pm$0.07 & 3.94$\pm$0.06 & $110^{+7}_{-7}$ &  $1.41^{+0.20}_{-0.17}$ & 100 & 1.27   & 0.04 &  2.3 \\
P6    & +0.223 & --0.512 & 2750 &   2.9 & +1.70$\pm$0.04 & 3.82$\pm$0.07 &   $90^{+5}_{-5}$ &  $0.83^{+0.07}_{-0.07}$ & 100 & 0.75   & 0.05  &  2.9 \\
                   \multicolumn{13}{c}{\textbf{Western disk side}}       \\
P7    & --0.133 & +0.036 &   250 & 34.2 & --0.56$\pm$0.03 & 7.12$\pm$0.13 & $279^{+2}_{-2}$     &$40.92^{+0.43}_{-0.37}$ &  70 & 6.53 & 0.17 &11.1 \\
P8    & --0.193 & +0.127 &   750 & 14.7 & --0.20$\pm$0.03 & 5.98$\pm$0.10 & $198^{+3}_{-3}$     &$11.01^{+0.27}_{-0.27}$ &  95 & 4.00 & 0.11 &  6.6 \\
P9    & --0.252 & +0.218 & 1250 &   6.7 & +0.07$\pm$0.03 & 5.82$\pm$0.17 & $153^{+3}_{-3}$     &  $8.31^{+0.20}_{-0.28}$ &  95 & 4.33 & 0.06 &  3.9 \\
P10  & --0.311 & +0.310 & 1750 &   4.8 & +1.07$\pm$0.05 & 4.54$\pm$0.07 & $129^{+6}_{-5}$     &  $3.39^{+0.30}_{-0.25}$ &  90 & 2.53 & 0.05  &  3.3 \\
P11  & --0.371 & +0.401 & 2250 &   3.6 & +1.56$\pm$0.07 & 3.54$\pm$0.10 & $120^{+10}_{-10}$ &  $0.87^{+0.18}_{-0.23}$ & 100 & 0.86 & 0.04  &  2.7 \\
P12  & --0.430 & +0.492 & 2750 &   2.8 & +1.88$\pm$0.06 & 2.72$\pm$0.11 & $100^{+10}_{-10}$ &  $0.25^{+0.03}_{-0.02}$ &  65 & 0.33 & 0.04  &  2.6 \\
                   \multicolumn{13}{c}{\textbf{Northeastern apparent spiral\tablefootmark{b}}}           \\
P13  & +0.293 & +0.003 & 1822 &   3.0 & +1.33$\pm$0.05 & 6.55$\pm$0.16 & $119^{+4}_{-3}$ & $4.26^{+0.26}_{-0.28}$ & 114 & 2.28 & 0.04 & 2.3 \\
P14  & +0.408 & --0.002 & 2349 &   3.2 & +1.22$\pm$0.04 & 5.43$\pm$0.11 & $119^{+4}_{-4}$ & $3.55^{+0.22}_{-0.23}$ & 114 & 2.28 & 0.04 & 2.4 \\
P15  & +0.468 & --0.100 & 2656 &   3.0 & +1.15$\pm$0.06 & 6.78$\pm$0.10 & $116^{+5}_{-4}$ & $2.69^{+0.23}_{-0.17}$ & 114 & 1.40 & 0.04 & 2.3 \\
P16  & +0.445 & --0.215 & 2689 &   3.6 & +0.57$\pm$0.07 & 6.66$\pm$0.14 & $107^{+5}_{-4}$ & $3.89^{+0.39}_{-0.33}$ &  90 & 2.09 & 0.05 & 3.1 \\
P17  & +0.380 & --0.310 & 2613 &   3.6 & +1.24$\pm$0.06 & 4.69$\pm$0.15 & $102^{+4}_{-3}$ & $4.27^{+0.37}_{-0.32}$ &  90 & 3.25 & 0.05 & 3.2 \\
                   \multicolumn{13}{c}{\textbf{Southwestern apparent spiral\tablefootmark{b}}}           \\
P18  & --0.754 & --0.110 & 3020 & $<$\,0.6\tablefootmark{a} & +0.57$\pm$0.08 &  8.58$\pm$0.15 & $115^{+6}_{-6}$ & $2.29^{+0.23}_{-0.17}$ &  90 & 0.94 & 0.01 & 0.5 \\
P19  & --0.800 & --0.007 & 3196 & $<$\,0.6\tablefootmark{a} & +1.56$\pm$0.09 & 11.51$\pm$0.18 & $110^{+15}_{-12}$ & $0.69^{+0.26}_{-0.12}$ &  90 & 0.21 & 0.01 & 0.5 \\
P20  & --0.752 & +0.098 & 3017 & $<$\,0.6\tablefootmark{a} & +0.81$\pm$0.09 &  6.95$\pm$0.13 & $90^{+15}_{-15}$ & $0.32^{+0.34}_{-0.12}$ &  90 & 0.16 & 0.01  & 0.6 \\
P21  & --0.651 & +0.152 & 2620 & $<$\,0.6\tablefootmark{a} & +0.77$\pm$0.10 &  6.66$\pm$0.08 & $110^{+10}_{-13}$ & $0.60^{+0.20}_{-0.11}$ &  90 & 0.32 & 0.01  & 0.5 \\
P22  & --0.540 & +0.178 & 2181 &  0.8 & +0.35$\pm$0.08 & 5.90$\pm$0.08 & $121^{+10}_{-9}$ & $1.07^{+0.28}_{-0.22}$ &  90 & 0.63 & 0.01 & 0.6 \\

\hline
\end{tabular}
\tablefoot{Column\,1: position label with reference to Fig.\,\ref{fig1}. Columns\,2 and\,3: offset positions of each pointing with respect to the
phase center. Column\,4: projected distance from the star position. Column\,5: integrated continuum flux at 860\,$\mu$m. \emph{Following parameters in columns 
6 to 11 were estimated by modeling the $19_{K}$\,$\to$\,$18_{K}$ emission of CH$_3$CN and CH$_3^{13}$CN with Weeds and MCWeeds assuming LTE conditions,
and they are shown with their statistical uncertainties.} Columns\,6 and~7: line velocity offset, with respect to 77.4\,km\,s$^{-1}$, and intrinsic line FWHM.
Columns\,8 to 10: rotational temperature and column density of molecular gas and source size. Column\,11: opacity of the CH$_3$CN\,($19_{4}$--$18_{4}$)
line. Columns\,12 and~13: gas mass and H$_2$ column density estimated from S$_{\rm dust}$ within a circular area of 500\,au diameter each. \tablefoottext{a}{Upper
limit set at 3\,$\sigma$.} \tablefoottext{b}{The northeastern and southwestern directions are named here with respect to the disk midplane 
as in Fig.\,\ref{fig2}.}}
\end{table*}

\section{Results}\label{res}

In Fig.\,\ref{fig1}, we present the brightness map of a CH$_3$OH line at 349.107\,GHz with an excitation energy of 260\,K, which is the brightest line emission 
in our band. The emission was integrated in velocity around ($\pm$2\,km\,s$^{-1}$) the line peak to emphasize the gas spatial distribution; this emission extends
within a radius of approximately 3000\,au from the young massive star in the region (star symbol) at the current sensitivity. For comparison with our cycle-3 ALMA
observations, we draw the disk midplane at three radii as marked in Fig.\,1 of \citet{Sanna2019}, from the central star position to 1000\,au (yellow), from 1000 to
2000\,au (black), and up to 3000\,au (magenta).  In Fig.\,\ref{fig2}, we plot the same image rotated clockwise by $57\degr$, in order to align the (projected) outflow 
direction with the vertical axis of the plot. The CH$_3$OH emission is centrally peaked near (but offset from) the star, and outlines two apparent spiral features which
extend on each side of the disk midplane. In Fig.\,\ref{figA1}, we also show that the CH$_3$CN gas has a similar spatial distribution at a similar excitation temperature
(for K\,$=$\,4). In Fig.\,\ref{figA2}, we present a continuum map of the dust emission at the center frequency of 349.150\,GHz, where the contours of Fig.\,\ref{fig1}
have been superposed for comparison. 

We have extracted the integrated spectra at 22 distinct positions in the frequency range 348.7 to 349.5\,GHz; this range includes the K-ladder of the 19\,$\to$\,18
rotational transition of the CH$_3$CN molecule and its isotopologue CH$_3^{13}$CN  (Table\,\ref{tlines}). Each position covers a circular area of radius 250\,au 
and they are plotted and labeled with white numbers in Fig.\,\ref{fig1}; these positions are used to sample the disk midplane and its apparent spiral features and, hereafter,
are referred to as P1, P2,..., P22. Examples of the CH$_3$CN spectra are shown and analyzed in Fig.\,\ref{fig2}, while the complete set of 22 spectra is reported
in Figs.\,\ref{figA3} and \ref{figA4}. We analyzed these spectra in two steps.

In step-1, we made use of the Weeds package of GILDAS \citep{Maret2011} to reproduce the CH$_3$CN  and CH$_3^{13}$CN spectra and estimate the
physical conditions of the emitting gas. Under the assumption of local thermodynamic equilibrium (LTE), and accounting for the continuum level,
Weeds produces a synthetic spectrum of an interstellar molecular species, depending on five parameters: \textbf{(i)} the intrinsic full width at half maximum (FWHM)
of the spectral lines;  \textbf{(ii)} the line offset with respect to the rest velocity in the region; \textbf{(iii)} the spatial extent (FWHM) of the emitting region, assumed
to have a Gaussian brightness profile; \textbf{(iv)} the rotational temperature and \textbf{(v)} column density of the emitting gas. We fixed parameters \textbf{(i)} and
\textbf{(ii)} by fitting a Gaussian profile to five K-components simultaneously, which are forced to have same linewidth and whose separation in frequency is set to the
laboratory values (command MINIMIZE of CLASS). This procedure assumes that each spectral component is excited within the same parcel of gas. This ``observed''
linewidth was used to derive the ``intrinsic'' linewidth by correcting iteratively for the opacity broadening, with the opacity computed by Weeds
\citep[e.g., Eq.\,4 of][]{Hacar2016}. Parameter \textbf{(iii)} was varied about the beam size at discrete steps of 5\,mas; emission at the same distance from the star
was forced to have same size (only within the midplane). This assumption was verified a posteriori based on the fit convergence (step-2). The same set of parameters
was used to compute the spectra of CH$_3$CN  and CH$_3^{13}$CN species, whose relative abundance is set to 30 \citep{Wilson1994,Sanna2014}.

In step-2, we input the initial parameters estimated with Weeds into MCWeeds \citep{Giannetti2017}, which implements a Bayesian statistical analysis for an automated
fit of the spectral lines. We made use of a Monte Carlo Markov chains method to minimize the difference between the observed and computed spectra and derive the
rotational temperature \textbf{(iv)} and column density \textbf{(v)} of the emitting gas with their statistical uncertainties. Parameters \textbf{(i)}, \textbf{(ii)}, and
\textbf{(iii)} were fixed from step-1. The following additional criteria apply: we fitted simultaneously optically thin and partially opaque spectral lines with opacity
lower than 5;  the CH$_3$CN K\,$=$\,0--3 components were excluded from the fit (except for P6 and P12), because their profiles show signs of either filtering or an
excess of warm envelope emission (e.g., side panels in Fig.\,\ref{fig2}; cf. Appendix~B of \citealt{Ahmadi2018}); the number of lines processed by MCWeeds is $\geq7$
for each pointing.


\begin{figure}
\centering
\includegraphics [angle= 0, scale= 0.6]{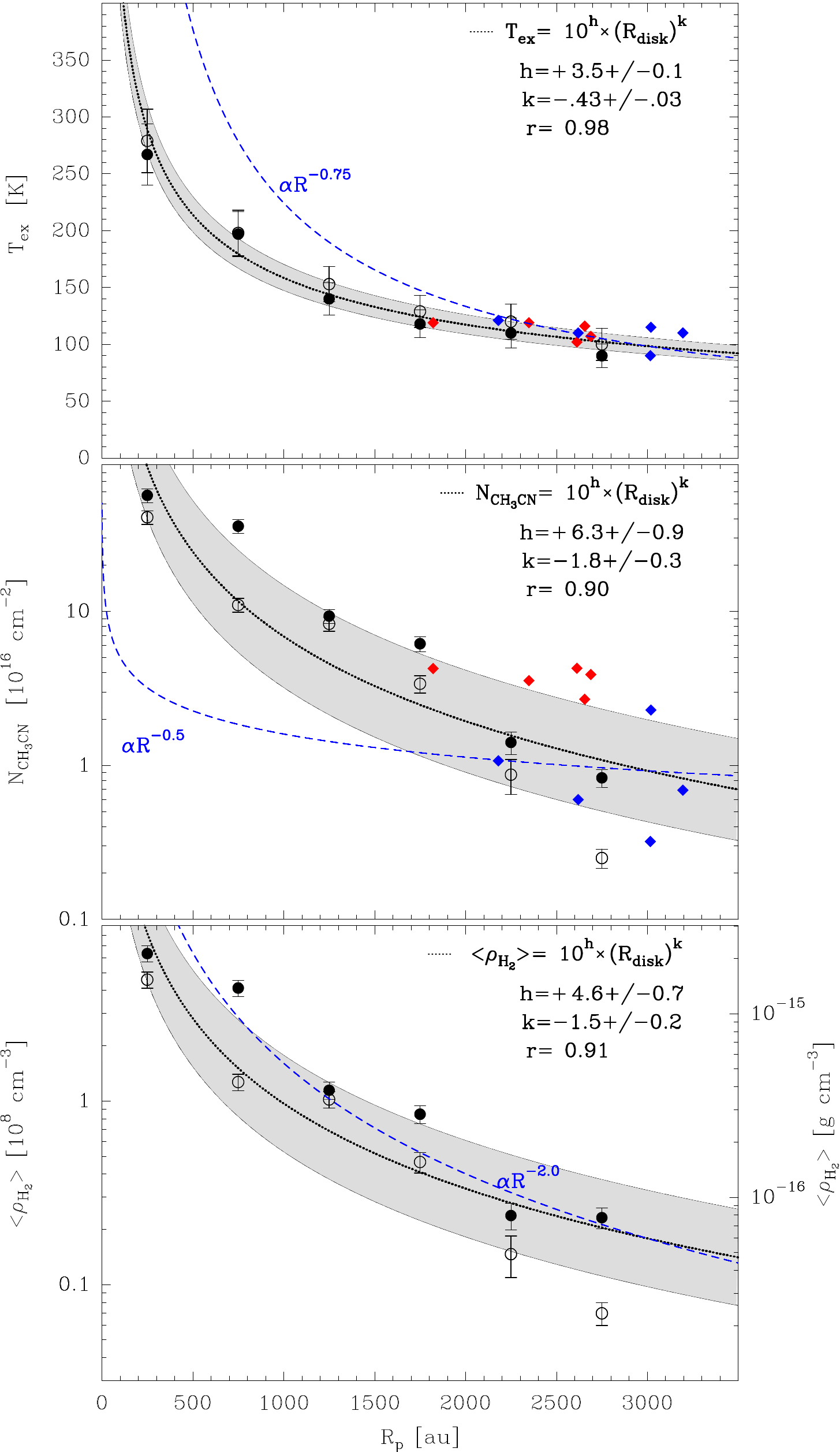}
\caption{Dependence of the gas physical conditions with the projected distance from the star, as determined from the relative intensities of the CH$_3$CN and
CH$_3^{13}$CN  lines near 349\,GHz. Filled and empty black circles mark measurements in the eastern and western disk sides, and red and blue diamonds mark
measurements in the northeastern ad southwestern apparent spirals, respectively (Table\,\ref{tpar}). Error bars of (only) midplane measurements are drawn, 
which account for an additional 10\% uncertainty due to the assumptions.
\textbf{Upper panel:} Temperature distribution along the disk midplane as a function of the projected distance from the star. The dotted bold line traces the $\chi^2$
fit to the sample distribution; the gray shadow shows the dispersion about the best fit  ($1\sigma$). The best fit parameters and the linear correlation coefficient are
indicated in the upper right. For comparison, the dashed blue line draws the increase of temperature with radius for a ``classical'' power-law dependence  (R$^{-0.75}$).
\textbf{Middle panel:} similar to the upper panel, but for the column density values. A logarithmic scale is used for the $y$-axis. 
\textbf{Lower panel:} similar to the upper panels, but for the H$_2$ volume density averaged along the line-of-sight; this is calculated by assuming an edge-on disk geometry.
The right axis provides the conversion to g\,cm$^{-3}$.}\label{fig3}
\end{figure}


Notably, in more than half of the 22 spectra of Figs.\,\ref{figA3} and \ref{figA4}, the methanol line at the center of the band stands brighter than the CH$_3$CN lines
at low excitation energies. Their emission is optically thick and sets an upper limit to the expected maximum brightness from gas in LTE. Assuming that the
CH$_3$OH and CH$_3$CN molecular species are fully coupled and emit cospatially, this is evidence that the CH$_3$OH\,($14_{1,13}-14_{0,14}$)\,A$^+$ 
transition is emitting by maser excitation, which adds to a plateau of thermal emission. In this context, we remind that G023.01$-$00.41 is among the brightest 
Galactic CH$_3$OH maser sources at 6.7\,GHz, and, for comparison, in Fig.\,\ref{figA5} we overplot to Fig.\,\ref{fig1} the positions of the 6.7\,GHz maser cloudlets
derived at milliarcsecond accuracy by \citet{Sanna2010,Sanna2015}. The distributions of the emission in different  CH$_3$OH maser lines usually resemble each other 
closely, and this evidence is confirmed in Fig.\,\ref{figA5} where, for instance, the brightest emission in both transitions clusters on the western side of the disk.
Also, we note that the same 349.1\,GHz CH$_3$OH transition was found to emit by maser excitation around the early-type young star, S255\,NIRS3 \citep{Zinchenko2018}.

In Table\,\ref{tpar}, we summarize the properties of the line and continuum emission associated with each of the 22 spectra: 
column~1 identifies each position as labeled in Fig.\,\ref{fig1}; 
columns~2  and~3 list the offset positions with respect to the phase center of the observations (Table\,\ref{tobs}); 
column~4 quantifies the (projected) linear distance of the center of each pointing from the star position, as defined in \citet{Sanna2019}; 
column~5 lists the integrated continuum flux at 860\,$\mu$m associated with each pointing, as measured from Fig.\,\ref{figA2};  
columns~6 to 10 list the best fit parameters of the gas emission output by MCWeeds with their statistical uncertainties, with column~11
specifying the opacity of the CH$_3$CN K\,$=$\,4 line;
columns~12 and~13 list the gas mass and H$_2$ column density estimated from the dust continuum emission\footnote{For a source of given distance ($d$)
and dust continuum flux (S$_{dust}$), the gas mass is evaluated from the following formula \citep{Hildebrand1983}: 
$$M = \frac{d^2 \cdot S_{dust} \cdot R }{\kappa_{\nu} \cdot B_{\nu}(T_{dust})} $$ 
where $R$ is the gas-to-dust mass ratio, $B_{\nu}$ is the Planck function at a dust temperature of $T_{dust}$, and $\kappa_{\nu}$ is the dust absorption
coefficient.} within an area of 500\,au diameter. This calculation assumes that the dust emission is optically thin at an equilibrium temperature equal to that
listed in column~8, with a standard gas-to-dust mass ratio of 100, and a uniform dust absorption coefficient of 1.98\,cm\,$^2$\,g$^{-1}$ at the observing
frequency, obtained for thick ice mantles, densities of 10$^8$ cm$^{-3}$, and linearly interpolated from the values tabulated by \citet{Ossenkopf1994}.
The validity of this assumption will be commented in Sect.~\ref{discus_phy}.

Although the synthetic spectra reproduce the bulk of the emission very well (red profiles in Figs.\,\ref{figA3}, \ref{figA4}), discrepancies between
the observed and expected emission of the higher K lines exist at the inner disk radii, where molecular chemistry is rich and contamination from different
molecular species is expected (e.g., CH$_3$OCHO, cf. \citealt{Liu2020}). Also, towards these regions the line of sight crosses large portions of gas at different
distances from the star, and one can expect that the integrated emission will be the combination of different gas physical conditions. Then, the working hypothesis
we have done, that only a single temperature and density component exists, may not be fully satisfied, and local gradients of both temperature and density might
affect the observed line profiles. This second order effect is neglected in the following analysis and will contribute to the uncertainty in the assumptions.

\section{Discussion}\label{discus}

\subsection{Physical conditions}\label{discus_phy}

Having on hand the information about the gas physical conditions across the extent of the disk, in the following we want to study the dependence of the gas temperature,
column and volume densities with distance from the central star. In Fig.\,\ref{fig3}, we present a diagram for each of these three quantities plotted against the projected 
distance (listed in column~4 of Table~\ref{tpar}), where disk loci are marked with different symbols: filled and empty black circles are used for the eastern
and western disk sides, respectively, and red and blue diamonds for the northeastern and southwestern apparent spirals, respectively. We explicitly note that 
the uncertainties reported in Table~\ref{tpar} quantify statistical errors only, meaning that they hold under the assumptions used for the analysis. While the statistical
errors convey information on the quality of the assumptions, they do not consider the uncertainties inherent in the assumptions themselves, which can be of the order of
10\% on average. Error bars plotted in Fig.\,\ref{fig3} account for a nominal uncertainty of 10\%  summed in quadrature to the statistical errors.

In the upper panel of Fig.\,\ref{fig3}, we show that the local gas temperature and the distance from the star are related to each other by a power-law of exponent
$-0.43$\,$\pm$0.03, and this relation is isotropic and verified both within the disk midplane and the apparent spirals. The dotted bold line draws the best fit to the data points,
obtained by minimizing a linear relation between the logarithms of the gas temperature and radius, $\log_{10}(\rm T_{ex})=k\,{\cdot}\,\log_{10}(\rm R_{p})+h$.
The gray shadow marks the dispersion about the best fit ($1\sigma$). Values of $k$ and $h$ are reported in the plot with their uncertainties, together with the linear
correlation coefficient of the sample distribution ($r$\,$=$\,0.98). Only for the sake of comparison, we draw a ``classical'' temperature
dependence with radius as R$^{-3/4}$, which is expected for geometrically thin disks heated externally by the star or internally by viscosity (e.g., \citealt{Kenyon1987});
this curve starts with the same temperature (98.4\,K) of the best fit at 3000\,au (the estimated outer disk radius). 

In the middle panel of Fig.\,\ref{fig3}, we plot the CH$_3$CN column density, in units of 10$^{16}$\,cm$^{-2}$, with respect to the same distance scale. The
dotted bold line draws the best fit to the sample distribution, similar to the upper panel, and the best fit parameters are reported in the plot with their uncertainties. 
In comparison with the upper panel, the data points are still strongly correlated (0.90), but show a steeper slope with a relative dispersion three times larger. 
Column densities along the apparent spiral arms are also consistent with the power law measured along the midplane. At variance, the dashed blue line shows a
shallower increase scaling with R$^{-0.5}$, as it would be expected for a spherical distribution of gas (e.g., an envelope/core) with density profile proportional
to R$^{-1.5}$; this curve starts with the same column density (9.2\,$\times$\,10$^{15}$\,cm$^{-2}$) of the best fit at 3000\,au. 

In the lower panel of Fig.\,\ref{fig3}, we can evaluate the average volume density of H$_2$ gas along the line-of-sight, <$\rho_{H_2}$>, under the assumption that the
central star is surrounded by a circular gaseous disk observed approximately edge-on, that has a sharp cutoff at radii of 3000\,au (R$_{\rm max}$). At each of the 12
pointings in the midplane, we divide the CH$_3$CN column density value by the local disk length intercepted by the observer, 
$2\times(R_{\rm max}^2-R_{\rm p}^2)^{0.5}$, where $R_{\rm p}$ is the projected distance of each pointing from the star position (as listed in Table\,\ref{tpar}). In
the calculation, we assume a constant relative abundance of 10$^{-8}$ between CH$_3$CN and H$_2$ species for temperatures above 100\,K, for consistency with
previous work \citep[e.g.,][]{Johnston2020,Ahmadi2019}. The best fit parameters are reported in the plot with their uncertainties, with the volume density in units of
10$^{8}$\,cm$^{-3}$. The volume density fit implies an average of 1.8\,$\times$\,$10^7$\,cm$^{-3}$ at a radius of 3000\,au, which increases to
8.2\,$\times$\,$10^8$\,cm$^{-3}$ at ten times smaller (projected) distance from the star. 

The average volume density evaluated at the projected radius ($R_{\rm p}$) of 250\,au is a robust underestimate for the gas density in the inner disk regions. At
the corresponding positions (P1 and P7 in Fig.\,\ref{fig1}), the line-of-sight crosses a large range of disk radii, from the outer disk regions down to 250\,au. We can
still exploit this average volume density to obtain a better approximation for the peak density at 250\,au from the star. The average integral of the best fit
power-law, $\frac{1}{(a-b)}\int_{b}^{a} 10^{4.6}\,\times\,R^{-1.5} dR$, would equal a volume density of 8.2\,$\times$\,$10^8$\,cm$^{-3}$ by definition, when
evaluated from 3000 ($a$) to 250\,au ($b$). By solving this equation at the inner radius, one derives a peak density of 4.8\,$\times$\,$10^9$\,cm$^{-3}$ (or
1.6\,$\times$\,$10^{-14}$\,g\,cm$^{-3}$) which is approximately a factor of 6 higher than the average. 

In column 12 and 13 of Table\,\ref{tpar}, we have estimated the mass and column density of H$_2$ gas from the dust continuum fluxes at 860\,$\mu$m,
assuming the dust emission is optically thin. However, the dust emission is optically thick at these wavelengths (with dust opacity of the order of unity and higher),
as it can be inferred by comparing the low brightness temperature of the continuum (of only a few 10\,K at the peak) with respect to the molecular line estimate
($\approx$\,300\,K). As a consequence, the dust continuum fluxes are only representative of emission in the envelope outskirts and do not probe the inner disk
regions, at variance with the CH$_3$CN line emission at high excitation energy. 
 

\begin{figure}
\centering
\includegraphics [angle= 0, scale= 0.35]{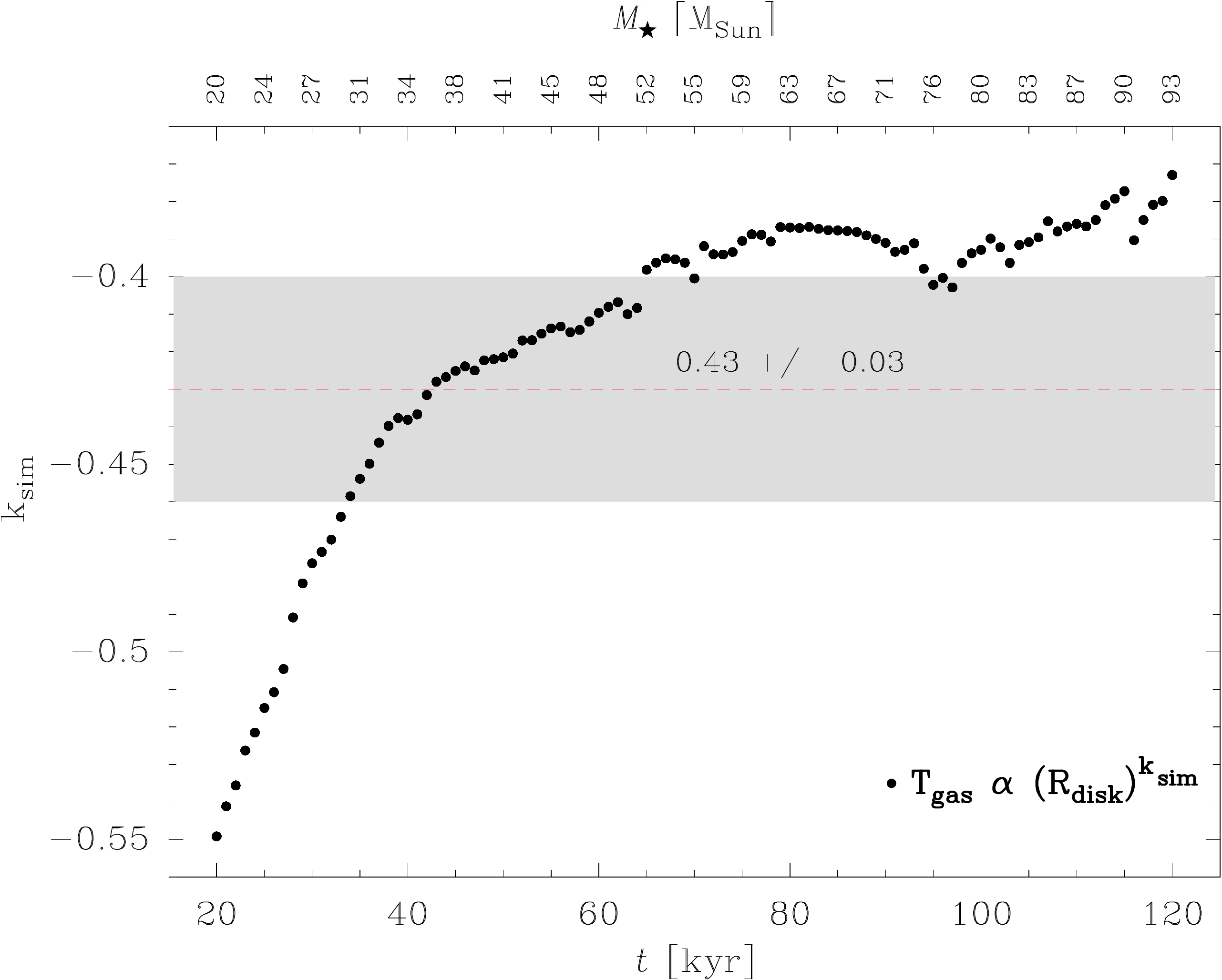}
\caption{Variation of the temperature slope with time, as determined at the disk midplane of a massive star with the simulations by
\citet[][their Fig.\,6]{Kuiper2018}, with the corresponding stellar mass indicated on top. k$_{sim}$ is the mean linear variation
determined over the range of radii, 100--2000\,au. The red dashed line marks the observed value fitted in the upper panel of Fig.\,\ref{fig3},
and the gray shadow defines the dispersion about the best fit ($1\sigma$).}\label{fig4}
\end{figure}


\subsubsection{Model comparison}

We can compare the above quantities with those derived in recent hydrodynamics simulations by \citet{Kuiper2018}, who simulated the growth of a massive star under 
the influence of its own radiation field and the effect of photoionization on the circumstellar medium. Their Fig.\,6 describes the evolution in time of the gas physical 
conditions along the disk midplane, as a function of the radius from the star. 

The average volume density in the lower panel of Fig.\,\ref{fig3} herein is diluted along the line-of-sight and it is systematically lower than the volume density expected at
an actual radius equal to the projected radius (R$_{\rm p}$). Therefore, we corrected the volume density at a projected radius of 250\,au and calculated a peak density
of 1.6\,$\times$\,$10^{-14}$\,g\,cm$^{-3}$ (see Sect.~\ref{discus_phy}): this estimate is in excellent agreement with the density predicted by the simulations at the
same radius, within a factor of 2. 

In Fig.\,\ref{fig4}, we present a plot of the temperature slope at the disk midplane (k$_{sim}$\,$=$\,$\frac{\partial \log_{10}(\rm T)}{\partial \log_{10}(\rm R)}$) as a
function of time, as derived from the simulations by \citet[][their Fig.\,6]{Kuiper2018}. The slope is defined by linear fitting the change of temperature with radius in the
distance range 100--2000\,au, and varies (almost) monotonically over a small range between $-0.55$ and $-0.37$. The axis on top indicates the corresponding 
stellar mass at a given time. The steeper slope at early times describes the settling of a very-young (with lower mass) accreting star and the formation of its
circumstellar disk. The observed slope is fully consistent with the model expectation and is drawn with a dashed line  in Fig.\,\ref{fig4} for comparison. The magnitude of
the modeled temperature at a given radius depends on the stellar mass. In Fig.\,6 of \citet{Kuiper2018}, the temperature at a radius of 250\,au increases with time from
300\,K to 500\,K, when the stellar mass approximately increases from 20\,M$_{\odot}$ to 100\,M$_{\odot}$. The lower end of this range is consistent with the
combination of stellar mass (20\,M$_{\odot}$) and disk temperature (289\,K) of our observations within the uncertainties, suggesting that the stellar system is still 
in a very early stage of evolution ($<$\,40\,kyr).

The comparison above shows that those basic quantities, obtained for the model and with our observations independently, are consistent with each other and 
validates these conditions for future model developments.

\subsection{Disk stability}

In the following, we want to study whether opposite forces are in equilibrium inside the accretion disk, or whether self-gravity might dominate locally causing the disk to
fragment. For this purpose, we apply the analysis by \citet{Kratter2010} who describe the stability of a disk where gas is falling in rapidly, as it is observed in the target
source. In \citet{Sanna2019}, we quantified a mass infall rate of $6\times10^{-4}$\,M$_{\odot}$\,yr$^{-1}$ by comparing position-velocity diagrams  observed along
the disk midplane with synthetic diagrams simulated through a dedicated disk model.

\begin{figure}
\centering
\includegraphics [angle= 0, scale= 0.35]{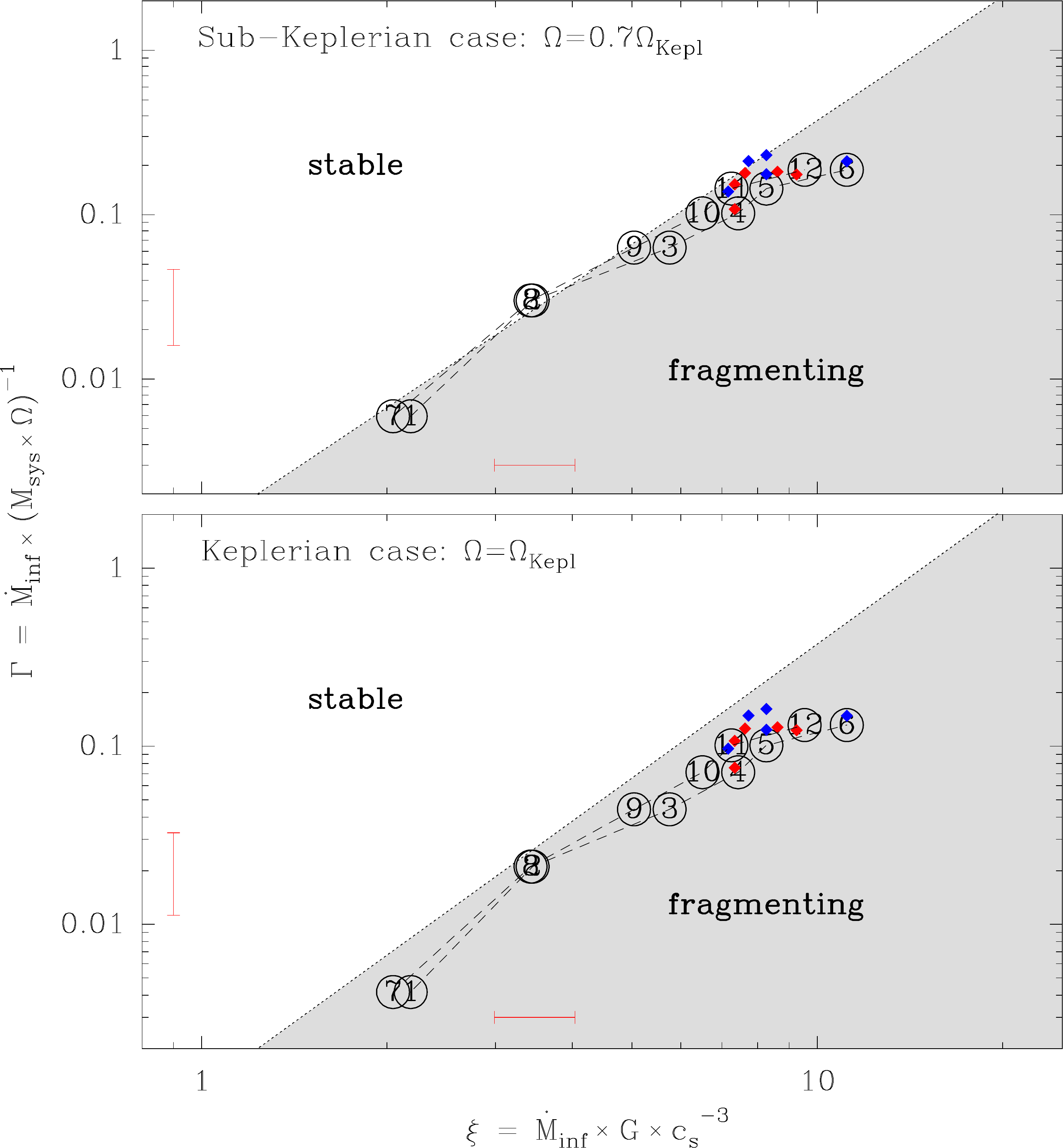}
\caption{Analysis of the disk stability against local fragmentation, following the methodology outlined in \citet{Kratter2010}. The plot reports the dependence
of two dimensionless parameters, $\Gamma$ and $\xi$, whose relation defines two regions where the disk is stable (white) or prone to fragmentation 
(grey). Black empty circles mark the local values of $\Gamma$ and $\xi$ within the midlplane and are labeled according to the pointing numbers in Table\,\ref{tpar}
(P2 and P8 overlap); red and blue diamonds mark values for the northeastern and southwestern apparent spirals, respectively. The upper and lower plots compare the
behavior for an angular velocity ($\Omega$) which is either 70\% of (upper), or equal to (lower), the Keplerian value. Reference error bars are indicated in red 
for an inner position (largest ticks in logarithmic scale), and account for a 10\% uncertainty in the local temperature and a positional uncertanty of half the
beam size.}\label{fig5}
\end{figure}

Following \citet{Kratter2010}, we have calculated two dimensionless accretion rates describing the state of the system. The first ``rotational'' parameter, 
$\Gamma$\,$=$\,$\rm \dot{M}_{inf}\cdot(M_{sys}\cdot\Omega)^{-1}$, relates the accretion timescale to the orbital timescale of infalling gas, where 
$\rm \dot{M}_{inf}$, $\rm M_{sys}$, and $\Omega$ are the mass accretion rate, the system mass of star plus disk at a given radius, and the angular velocity
at the same radius, respectively. The second ``thermal'' parameter, $\xi $\,$=$\,$\rm \dot{M}_{inf} \cdot G \cdot c_{s}^{-3}$, relates the mass accretion
rate to the local sound speed of disk material. $\rm \dot{M}_{inf}$ and M$_{sys}(R)$ were quantified in \citet{Sanna2019}; $\Omega(R)$ is evaluated
based on the enclosed mass at a given radius and considering an orbital motion which is either Keplerian or 70\% of its value; $c_{s}$ is evaluated from
the excitation temperatures in Table\,\ref{tpar}\footnote{The sound speed is calculated from the formula, $c_{s}$\,$=$\,$\sqrt{\rm \gamma k_B T / \rm \mu m_H}$,
where $\gamma$\,$=$\,$7/5$ is the adiabatic index for a diatomic gas (H$_2$), k$\rm _B$ is the Boltzmann constant,  $\mu$\,$=$\,$2.8$ is the mean 
molecular weight per hydrogen molecule, and m$\rm_H$ is the mass of a hydrogen atom.}.  These parameters are calculated for each of the 22 pointings marked
in Fig.\,\ref{fig1}, and are plotted in Fig.\,\ref{fig5} for a direct comparison with Fig.\,2 of \citet{Kratter2010}.  

In Fig.\,\ref{fig5}, the upper and lower panels describe the state of the system for a sub-Keplerian and Keplerian rotation curve, respectively. We have
demonstrated that the system is sub-Keplerian outside of 500\,au from the central star and might approach centrifugal equilibrium inside this radius
\citep{Sanna2019}. For this reason, we expect the system to behave in between these two extremes. The diagram is divided in two regions where the
system is either stable (white) or prone to fragmentation (grey); this empirical boundary condition is defined at $\Gamma$\,$=$\,$\rm \xi^{2.5} \cdot 850^{-1}$
\citep{Kratter2010}. $\Gamma-\xi$ couples characterizing the disk midplane are labeled according to the numbers in Fig.\,\ref{fig1}; $\Gamma-\xi$ couples
characterizing the apparent spiral features are grouped over a small range of values in the plot and are indicated with red and blue diamonds for clarity (same symbols
as in Fig.\,\ref{fig3}).

This analysis outlines that $\Gamma-\xi$ couples are distributed near the boundary between a stable and unstable disk, and the accretion disk meets the
condition for fragmentation marginally along its whole extent. Regions beyond 1000\,au are progressively more unstable and have the higher likelihood
to produce  stellar companions. The parameter $\xi$ depends on T$^{-3/2}$ through the sound speed, and explains the small shifts between the warmer western
side of the disk and the colder eastern side (by approximately 10\,K). The disk temperature increases with time by tens to hundreds of Kelvin due to the growing
stellar mass, and, as a consequence, thermal pressure would act to stabilize the disk whose position would shift to the left side of the diagram. Concurrently, 
the disk gains mass and increases its Keplerian radius as time passes \citep[][their Fig.\,7]{Kuiper2018}, so that the outer disk regions will accelerate from
sub-Keplerian (Fig.\,\ref{fig5}, upper panel) to Keplerian rotation (Fig.\,\ref{fig5}, lower panel) on average. The parameter $\Gamma$ decreases by increasing
the mass accreted by the star plus disk and its angular velocity: such a variation makes the disk shifts in the diagram to a less stable configuration (cf. Sect.\,4 of 
\citealt{Kratter2010} for the relation to the Toomre parameter). Whether or not the system will grow more unstable in time as an overall effect, it is difficult
to predict without further information about, for instance, the history of the system, which might have already developed internal fragments in the past.

From a theoretical point of view, and in agreement with our snapshot taken in Fig.\,\ref{fig5}, unstable accretion disks are a robust product of distinct models
simulating the formation of massive stars  \citep[e.g.,][]{Kratter2010,Klassen2016,Rosen2016,Harries2017,Meyer2017,Meyer2018,Rosen2019,Ahmadi2019,Oliva2020}.
Unstable disks can fragment leading to hierarchical star systems that consist of (several) low-mass stellar companions surrounding a massive primary. These
simulations also predict that unstable accretion disks will develop substructures and density-enhanced spiral features, which wrap along the equatorial plane. In the
following section, we comment on the apparent spiral morphology imaged in Fig.\,\ref{fig1}.


\subsubsection{Spiral arms or accretion streams?}

Spiral arms can be produced by gravitational instabilities, being the manifestation of underlying density waves in a self-gravitating disk \citep[e.g.,][]{Lodato2005},
and by tidal interactions between the disk and a cluster companion, either internal or external to the disk itself \citep[e.g.,][]{Zhu2015}. 

From an observational point of view, evidence for disk substructures have been recently reported towards a couple of massive young stars \citep{Maud2019,Johnston2020}.
In particular, \citet{Johnston2020} have outlined the existence of a spiral arm branching off the disk of a massive star at varying pitch angle (20$\degr$--47$\degr$).
Arguably, the asymmetric spiral morphology they found could be triggered by tidal interactions with cluster companions \citep[e.g.,][]{Forgan2018}.

In order to interpret the apparent spiral morphology imaged towards G023.01$-$00.41, we have to take into account the viewing angle to the observer.  
In Fig.\,\ref{fig2}, the horizontal axis of the plot approximately coincides with the equatorial plane of the disk, whose inclination was inferred to be nearly edge-on.   
Consequently, the vertical axis of the plot provides approximate altitudes above the disk midplane, implying that the apparent spiral features extend up to 
2500-3000\,au off the plane (up to 3500\,au in the limit of 30$\degr$ inclination). This evidence suggests that the apparent spiral features might be rather streams
of gas producing a warp in the outer disk regions and, hereafter, we refer to the northeastern and southwestern apparent spirals as the northeastern and southwestern
streams, respectively. Whether these streams of gas are due to inward or outward motions can be tested based on the line fitting presented in Sect.\,\ref{res}.
 
The combined fitting over a group of spectral lines, emitting from the same spatial region, allows us to determine the offset velocity (v$_{\rm off}$) of local
gas with an accuracy better than $0.1$\,km\,s$^{-1}$ (column\,6 of Table\,\ref{tpar}). This analysis reveals two velocity gradients of the order of 1\,km\,s$^{-1}$
per 2000\,au along the northeastern (positions 13 to 16) and southwestern (positions 19 to 22) streams. Gas accelerates from the ambient velocity
in the outer stream regions, at about 79\,km\,s$^{-1}$, to blueshifted velocities close to the disk plane. If disk material were moving outward, as due to a disk
wind, one would expect the gas velocity to be very different from the ambient velocity at the stream tip. On the  contrary, these velocity gradients are consistent with a
scenario where gas accelerates towards the disk moving from the outer envelope at the ambient velocity, and supporting  the existence of an accretion flow. This scenario
naturally explains the infall profile previously detected in the position-velocity diagrams along the disk midplane \citep[][their Fig.\,5]{Sanna2019}. At positions P17 
and P18, the measured velocity differs from that expected for a regular gradient along the entire streams. This discrepancy can be interpreted as an effect of
contamination from the outer disk gas at position P17, and the local gas dynamics at position P18 (see further discussion below). For completeness, we also
mention a third scenario where the apparent spiral features might just be overdensities within the larger envelope around the disk, although we consider this hypothesis
very unlikely given the spatial morphology and physical conditions observed.

Explaining the three-dimensional morphology of the accretion streams prompts for a dedicated theoretical ground which goes beyond the scope of the current paper.
Therefore, in the following we highlight a number of observational features which should be taken into account in future dedicated simulations. First, the northeastern
and southwestern streams appear approximately symmetric in gas emission but not in dust, with the southwestern stream being deficient in continuum emission with
respect to the northeastern stream (Fig.\,\ref{figA2}). Whether this difference is related to the origin of the streams themselves should be clarified. Second, to the east
of the accretion disk, at a (projected) distance of approximately 3500\,au from the star, the continuum map shows a dust overdensity which also coincides with a local
peak of molecular emission, labeled as ``mm2'' in  Fig.\,\ref{fig2}. Whether and how this source could perturb the stream morphology and affect the local warp of the
disk should be clarified. Finally, we note that the outer tips of both streams are the regions with the higher combination of column density and temperature along the
streams (P13 and P18 of Table\,\ref{tpar}). Simulations of in-plane spiral arms show that their overdensities can fragment leading to companion stars that, in turn, can
influence the morphology of disks and spirals \citep[e.g., Fig.\,1 and 12 of][]{Oliva2020}. By analogy, we pose the question of whether or not the stream tips could host
newly formed stars, and, if yes, how they could interact with the stream morphology.

\section{Conclusions}\label{concl}

We report on spectroscopic Atacama Large Millimeter/submillimeter Array (ALMA) observations near 349\,GHz with a spectral and angular resolutions of
0.8\,km\,s$^{-1}$ and $0\farcs1$, respectively. We targeted the star-forming region G023.01$-$00.41 and imaged the accretion disk around a luminous young
star of $10^{4.6}$\,L$_{\odot}$ in both methyl cyanide and methanol emission. We have fitted the K-ladder of the CH$_3$CN\,($19_{K}$--$18_{K}$) transitions, 
and that of the isotopologue CH$_3^{13}$CN, to derive the physical conditions of dense and hot gas within 3000\,au from the young massive star. 
Our results can be summarized as follows:

\begin{enumerate}

\item We have resolved the spatial morphology of the accretion disk which shows two apparent spiral arms in the disk outskirts at opposite sides (Fig.\,\ref{fig2}, main).
These apparent spirals likely represent streams of accretion from the outer envelope onto the disk. They appear almost symmetric in molecular gas but not in dust, with
one stream being deficient in continuum emission with respect to the other (Fig.\,\ref{figA2}). The disk is remarkably bright in CH$_3$OH\,($14_{1,13}-14_{0,14}$)\,A$^+$
emission showing a maser contribution up to (at least) 2000\,au from the young star. 

\item We have derived the temperature, column and volume densities of molecular gas along the disk midplane and accretion streams, and studied their dependence
with distance from the central star (Fig.\,\ref{fig3}). The gas temperature varies as a power law of exponent R$^{-0.43}$, which is significantly shallower than that
expected around Solar-mass stars (R$^{-0.75}$). The volume density averaged along the line-of-sight through the disk varies as a power law of exponent R$^{-1.5}$;
this slope implies a peak density of 1.6\,$\times$\,$10^{-14}$\,g\,cm$^{-3}$ at 250\,au from the star.

\item Our findings are in excellent agreement with the results of recent hydrodynamics simulations of massive star formation \citep{Kuiper2018}, and this
comparison supports the idea we are tracing disk conditions as opposed to envelope conditions. In turn, this comparison provides evidence that the high
excitation-energy transitions of CH$_3$CN allow us to peer into the inner disk regions.

\item We have studied whether the disk is stable against local gravitational collapse following the analysis by \citet{Kratter2010}, who describe the stability of a disk
undergoing rapid accretion. This analysis outlines that the accretion disk marginally meets the condition for fragmentation along its whole extent, and regions beyond radii
of 1000\,au are progressively more unstable and have the higher likelihood to produce stellar companions (Fig.\,\ref{fig5}). Notably, the tips of the accretion streams are
the loci of higher temperature and column density, and could be hosting young stars in the making. The fact that the disk is gravitationally unstable, together with the
effect of tidal interactions with cluster member(s), might be the origin of disk substructures to be probed by observations of optically-thin dust emission.

\end{enumerate}

These observations provide direct constraints for models trying to reproduce the formation of stars of tens of Solar masses, with particular focus on the 
disk properties, the spatial morphology of gas accreting onto the disk, and the perspective of forming a tight cluster of stellar companions. 

\begin{acknowledgements}

Comments from an anonymous referee, which helped improving our paper, are gratefully acknowledged.
This paper makes use of the following ALMA data: ADS/JAO.ALMA$\#$2016.1.01200.S. ALMA is a partnership of ESO (representing its member states),
NSF (USA) and NINS (Japan), together with NRC (Canada), MOST and ASIAA (Taiwan), and KASI (Republic of Korea), in cooperation with the Republic of Chile.
The Joint ALMA Observatory is operated by ESO, AUI/NRAO and NAOJ. 
ACG acknowledges funding from the European Research Council under Advanced Grant No. 743029, Ejection, Accretion Structures in YSOs (EASY).
MB acknowledges financial support from the French State in the framework of the IdEx Universit\'{e} de Bordeaux Investments for the future Program.
RK acknowledges financial support via the Emmy Noether Research Group on Accretion Flows and Feedback in Realistic Models of Massive Star Formation funded by the German Research Foundation (DFG) under grant no. KU 2849/3-1 and KU 2849/3-2.

\end{acknowledgements}


\bibliographystyle{aa}
\bibliography{asanna1120}



\begin{appendix}


\section{Additional material}\label{Append}


\begin{table*}
\caption{Observed lines\label{tlines}}
\centering
\begin{tabular}{c c r}

\hline \hline
\multicolumn{1}{c}{$\nu$}  &  \multicolumn{1}{c}{Species\,/\,Line} & \multicolumn{1}{c}{E$_{\rm up}$}  \\
\multicolumn{1}{c}{(GHz)}  &                                         & \multicolumn{1}{c}{(K)}                  \\ 
\hline

 & & \\

                  \multicolumn{3}{c}{\textbf{CH$_3$CN}}              \\
348.784 & CH$_3$CN\,($19_{10}-18_{10}$)    &  880.6  \\
348.911 & CH$_3$CN\,($19_9-18_9$)    &  745.4  \\
349.024 & CH$_3$CN\,($19_8-18_8$)    &  624.3  \\
349.125 & CH$_3$CN\,($19_7-18_7$)    &  517.4  \\
349.212 & CH$_3$CN\,($19_6-18_6$)    &  424.7  \\
349.286 & CH$_3$CN\,($19_5-18_5$)    &  346.2  \\
349.346 & CH$_3$CN\,($19_4-18_4$)    &  281.9  \\
349.393 & CH$_3$CN\,($19_3-18_3$)    &  232.0  \\
349.426 & CH$_3$CN\,($19_2-18_2$)    &  196.3  \\
349.446 & CH$_3$CN\,($19_1-18_1$)    &  174.8  \\ 
349.453 & CH$_3$CN\,($19_0-18_0$)    &  167.7  \\ 
                  \multicolumn{3}{c}{\textbf{CH$_3^{13}$CN}}              \\
348.853 & CH$_3^{13}$CN\,($19_8-18_8$)    &  624.3  \\
348.953 & CH$_3^{13}$CN\,($19_7-18_7$)    &  517.3  \\
349.040 & CH$_3^{13}$CN\,($19_6-18_6$)    &  424.6  \\
349.113 & CH$_3^{13}$CN\,($19_5-18_5$)    &  346.1  \\
349.173 & CH$_3^{13}$CN\,($19_4-18_4$)    &  281.9  \\
349.220 & CH$_3^{13}$CN\,($19_3-18_3$)    &  231.9  \\
349.254 & CH$_3^{13}$CN\,($19_2-18_2$)    &  196.2  \\
349.274 & CH$_3^{13}$CN\,($19_1-18_1$)    &  174.8  \\
349.280 & CH$_3^{13}$CN\,($19_0-18_0$)    &  167.7  \\                   
                  \multicolumn{3}{c}{\textbf{CH$_3$OH}}              \\
349.107 & CH$_3$OH\,($14_{1,13}-14_{0,14}$)\,A$^+$   & 260.2 \\

\hline
\end{tabular}

\tablefoot{Frequencies and upper energy levels for each molecular transition are obtained from the CDMS \citep{Endres2016} and JPL \citep{Pearson2010} catalogues.}\\

\end{table*}




\begin{figure*}
\centering
\includegraphics [angle= 0, scale= 1.0]{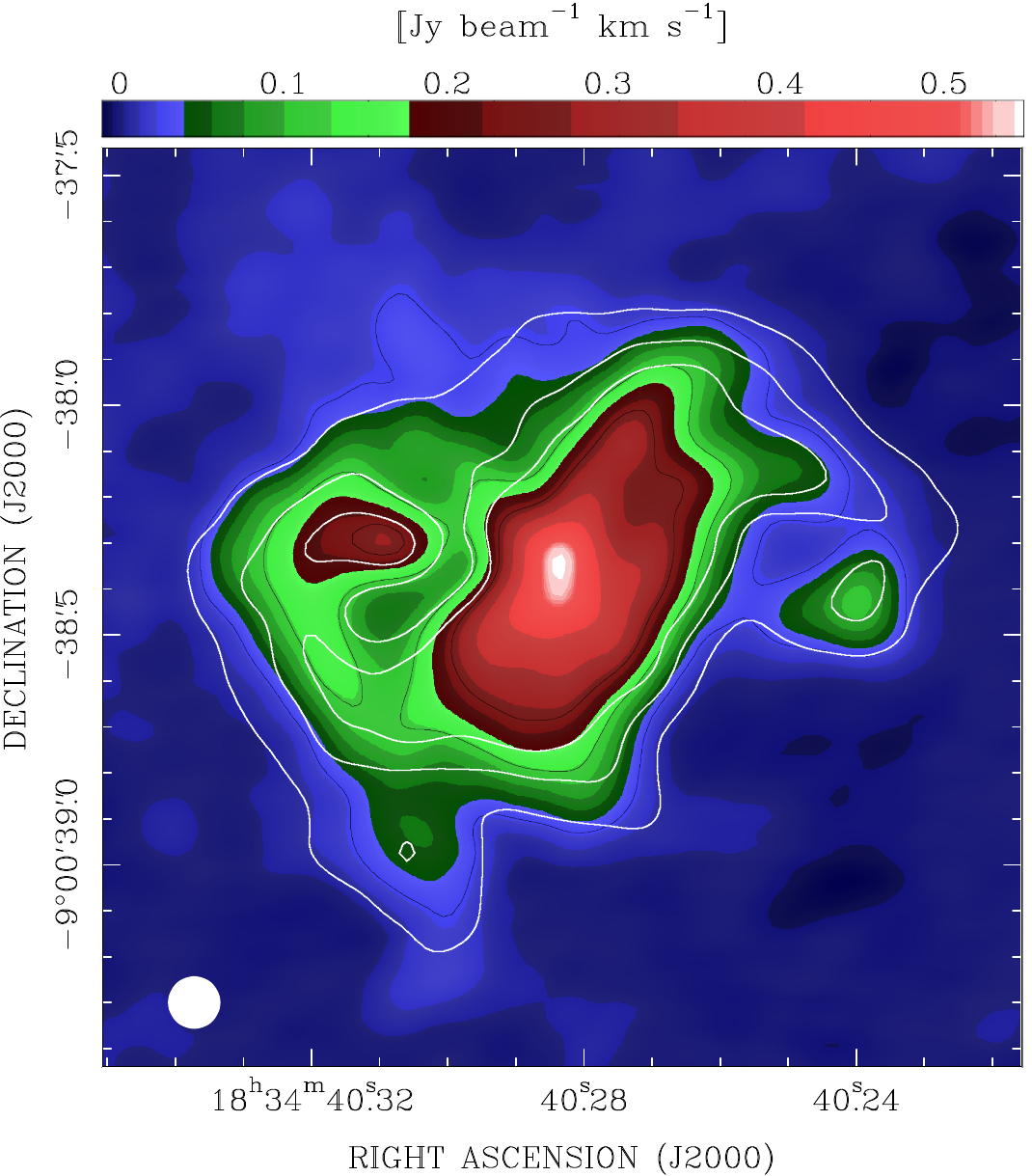}
\caption{Moment-zero map of the CH$_3$CN\,($19_{4}$--$18_{4}$) emission (colors and black contours) for a direct comparison with the
CH$_3$OH\,($14_{1,13}$--$14_{0,14}$)\,A$^+$ emission of Fig.\,\ref{fig1} (white contours). These CH$_3$CN and CH$_3$OH transitions have similar excitation
energies (E$_{\rm up}$) of 281.9\,K and 260.2\,K, respectively, and the line emission has been integrated over the same velocity range. The upper wedge quantifies 
the line intensity from its peak to the maximum negative in the map; black contours are drawn at levels of 10, 50, and 100 times the 1\,$\sigma$ rms of
2.5\,mJy\,beam$^{-1}$\,km\,s$^{-1}$. We note that the white and black contours should be used for a direct comparison between the spatial morphology of
the CH$_3$OH and CH$_3$CN emissions, respectively, because they correspond to relative levels of comparable rms (at variance with the color map which depends
on the maximum in the map). The synthesized ALMA beam is shown in the bottom left corner.}\label{figA1}
\end{figure*}



\begin{figure*}
\centering
\includegraphics [angle= 0, scale= 0.85]{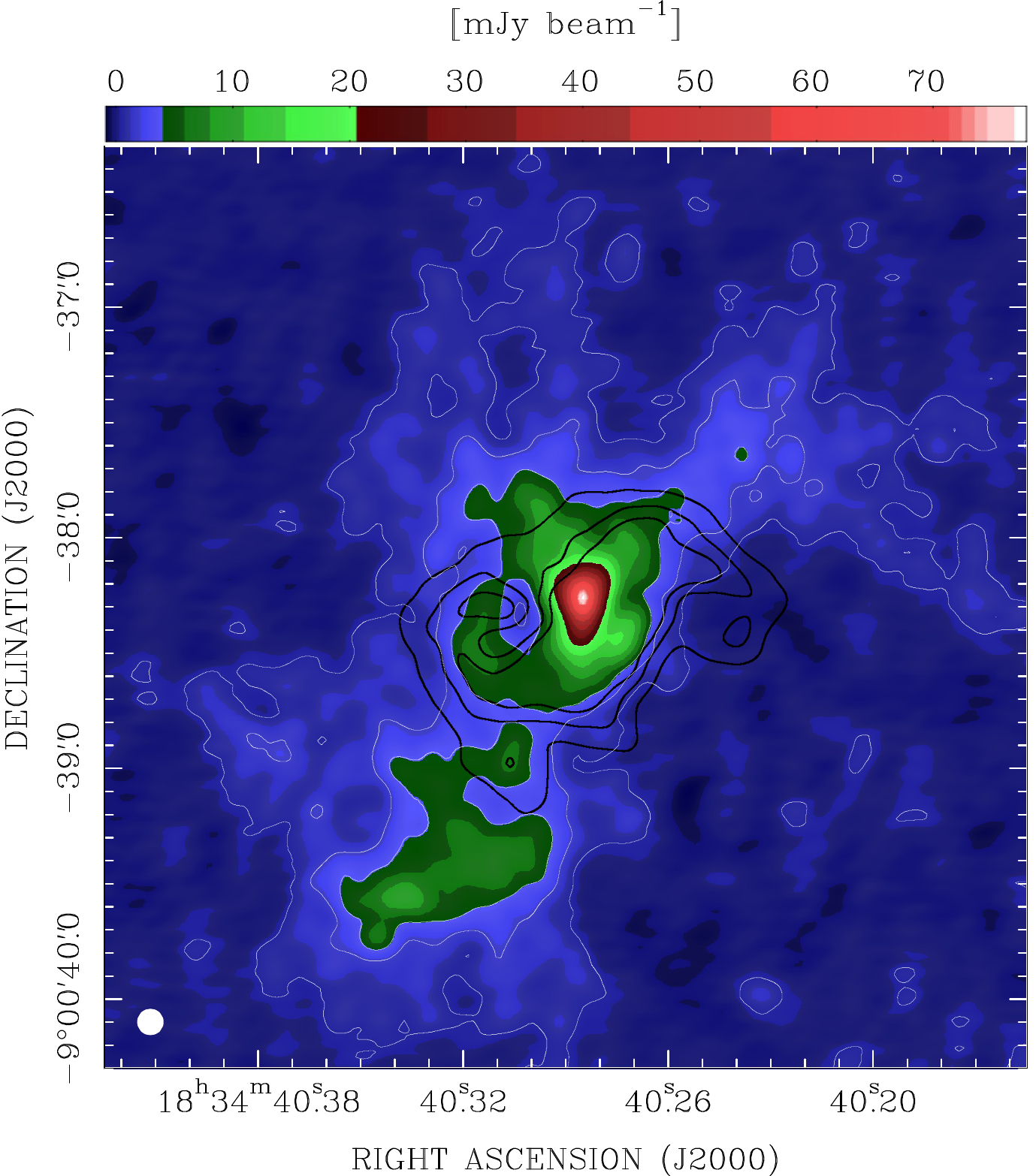}
\caption{Dust continuum emission at 860\,$\mu$m (colors and white contours) imaged with ALMA towards G023.01$-$00.41 and derived from 52\,MHz of 
emission-free band in the spectra. The upper wedge quantifies the continuum brightness from its peak to the maximum negative in the map; white contours are
drawn at levels of 3, 10, and 20 times the 1\,$\sigma$ rms of 0.2\,mJy\,beam$^{-1}$. The moment-zero map of the CH$_3$OH\,($14_{1,13}$--$14_{0,14}$)\,A$^+$ 
mission is overlaid with the same black contours of Figs.\,\ref{fig1} and \ref{fig2} for comparison. The synthesized ALMA beam is shown in the bottom left corner.}
\label{figA2}
\end{figure*}



\begin{figure*}
\centering
\includegraphics [angle= 0, scale= 0.81]{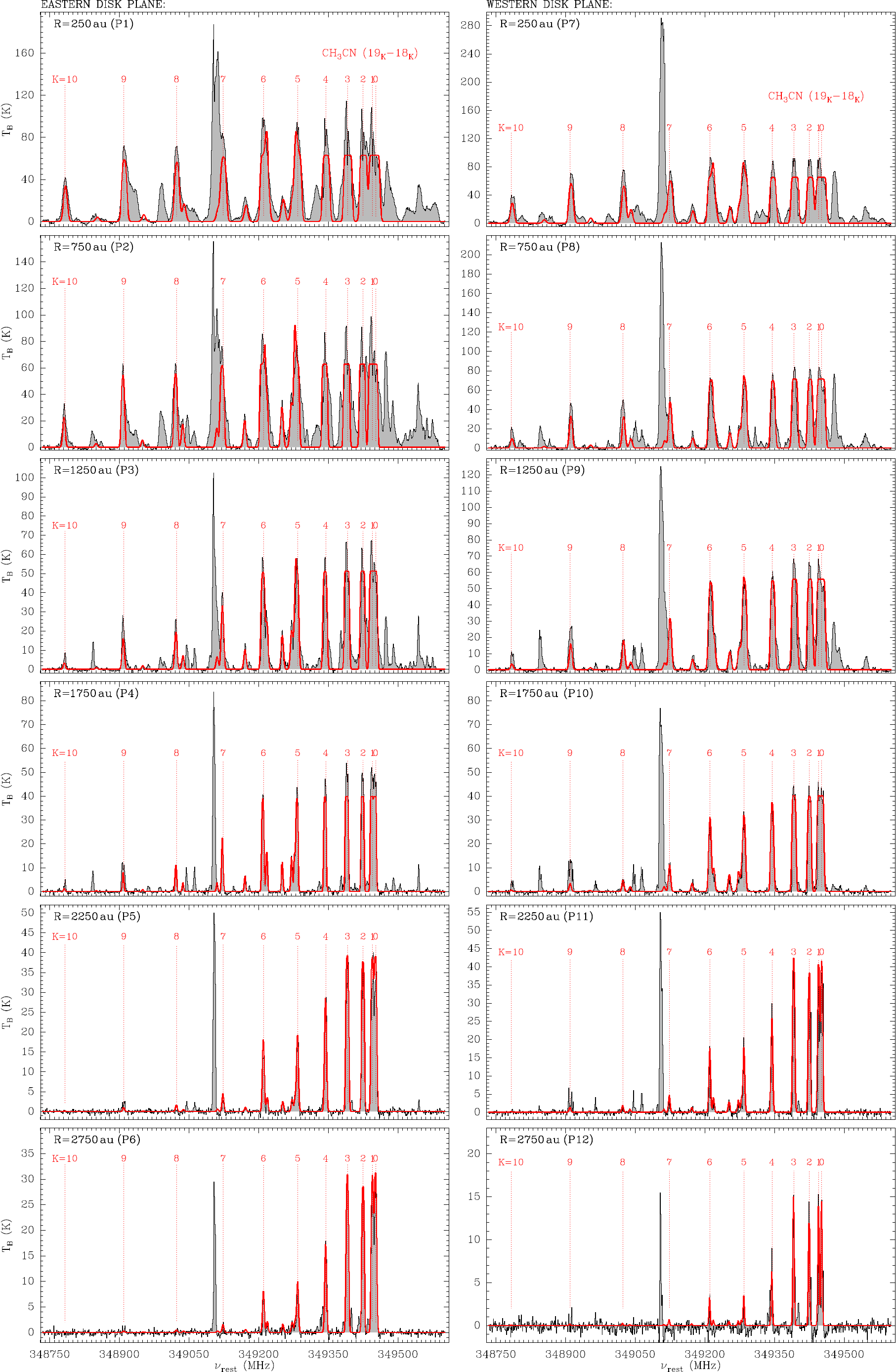}
\caption{Spectra of the CH$_3$CN\,($19_{K}$--$18_{K}$) transitions along the eastern (left column) and western (right column) sides 
of the disk plane. Spectra are integrated at radial steps of 500\,au as indicated in Fig.\,\ref{fig1}, and each pointing is labeled on the top left
accordingly. CH$_3$CN components, with K ranging from 0 to 10, are marked in red in each spectrum. The brightest line at a rest frequency of
349.107\,GHz corresponds to the CH$_3$OH\,($14_{1,13}$--$14_{1,14}$)\,A$^+$ emission imaged in Fig.\,\ref{fig2}. Modeled spectra
are overplotted in red; note that the first four K components were not fitted, except at positions P6 and P12 (cf. Sect.~\ref{res}).}\label{figA3}
\end{figure*}




\begin{figure*}
\centering
\includegraphics [angle= 0, scale= 1.00]{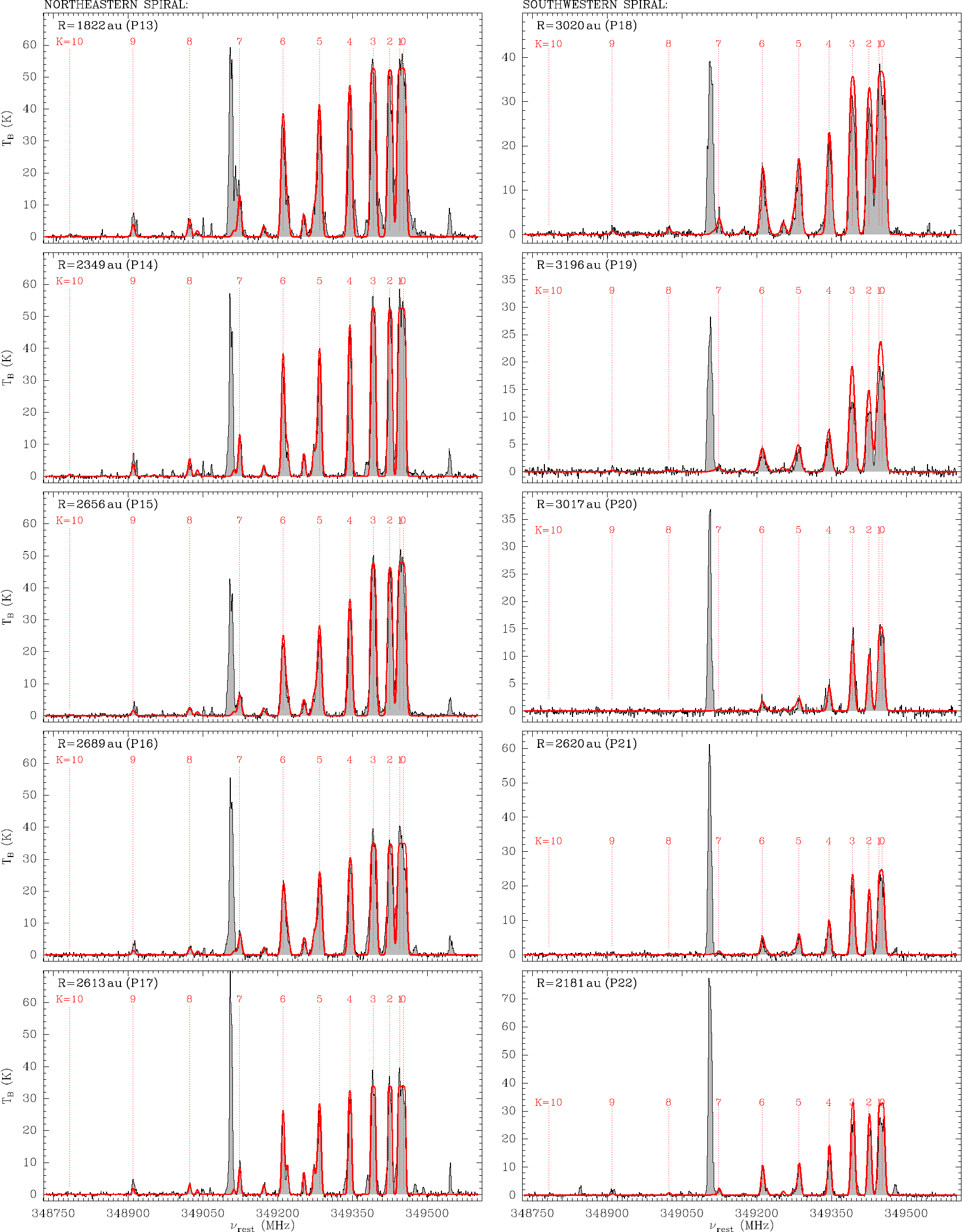}
\caption{Same as Fig.\,\ref{figA3}, but for spectra of the CH$_3$CN\,($19_{K}$--$18_{K}$) transitions along the northeastern (left column) and southwestern
(right column) spirals.}\label{figA4}
\end{figure*}




\begin{figure*}
\centering
\includegraphics [angle= 0, scale= 1.00]{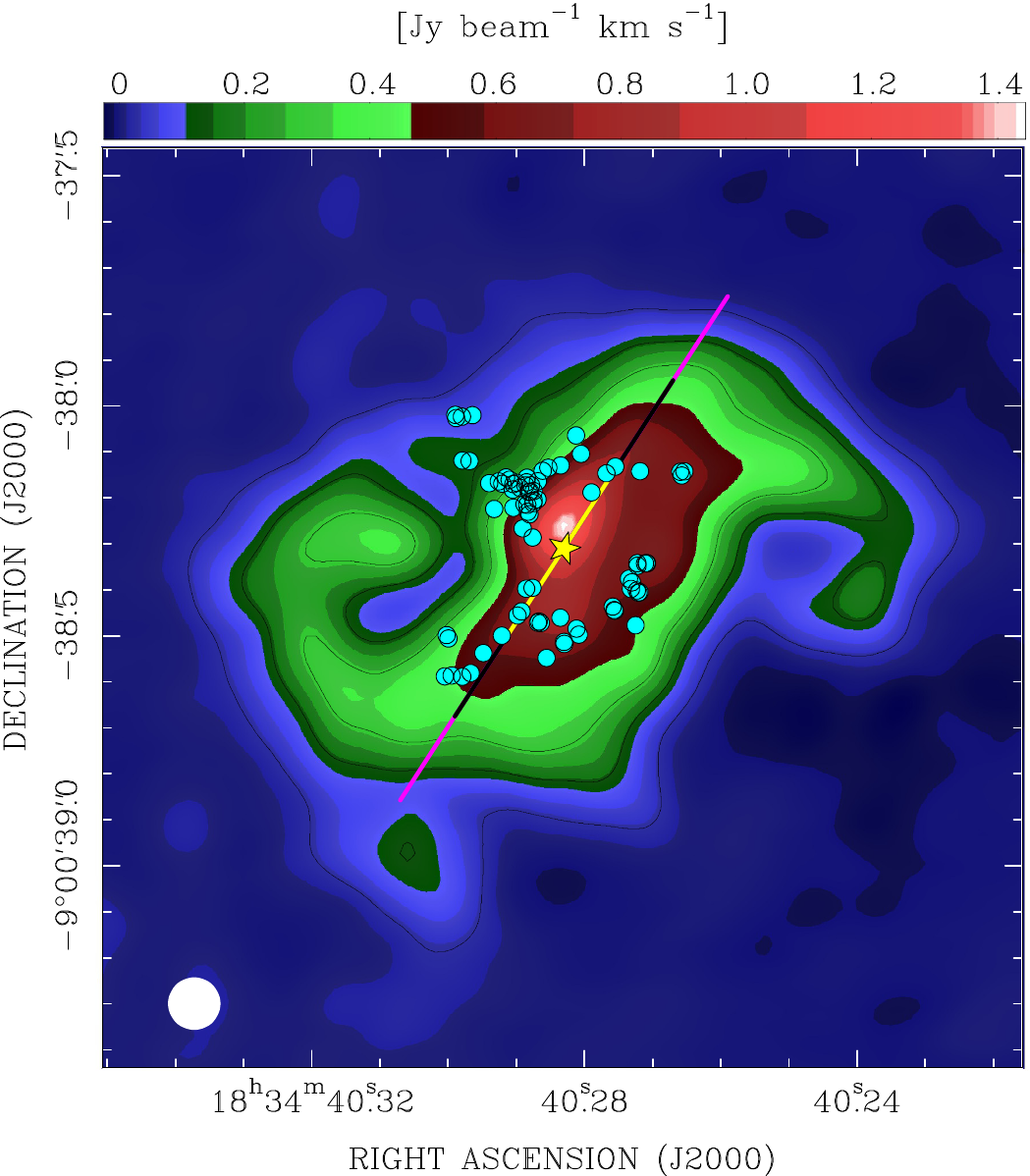}
\caption{Same as Fig.\,\ref{fig1}, but, for comparison, the positions of the 6.7\,GHz CH$_3$OH maser cloudlets are overplotted with cyan dots 
\citep[from][]{Sanna2010,Sanna2015}.}\label{figA5}
\end{figure*}


\end{appendix}

\end{document}